\documentclass[letterpaper,12pt]{article}   

\usepackage{rangecite}
\usepackage{osajnl} 

\usepackage[draft]{hyperref} 
\usepackage{color}

\usepackage{epsf}

\newcommand{\ct}{\chi^{(3)}}

\newcommand{\Mpn}{\bar{A}_{p}}
\newcommand{\Man}{\bar{A}_{a}}
\newcommand{\Msn}{\bar{A}_{s}}
\newcommand{\bpn}{{A}_{p}}
\newcommand{\ban}{{A}_{a}}
\newcommand{\bsn}{{A}_{s}}

\newcommand{\opn}{\hat{A}_{p}}
\newcommand{\oan}{\hat{A}_{a}}
\newcommand{\osn}{\hat{A}_{s}}
\newcommand{\opd}{\hat{A}_{p}^{\dagger}}
\newcommand{\oad}{\hat{A}_{a}^{\dagger}}
\newcommand{\osd}{\hat{A}_{s}^{\dagger}}

\newcommand{\fpn}{\hat{a}_{p}}
\newcommand{\fan}{\hat{a}_{a}}
\newcommand{\fsn}{\hat{a}_{s}}

\newcommand{\fpd}{\hat{a}_{p}^{\dagger}}
\newcommand{\fad}{\hat{a}_{a}^{\dagger}}
\newcommand{\fsd}{\hat{a}_{s}^{\dagger}}

\newcommand{\Mpd}{\bar{A}_{p}^{*}}
\newcommand{\Mad}{\bar{A}_{a}^{*}}
\newcommand{\Msd}{\bar{A}_{s}^{*}}

\newcommand{\bad}{{A}_{a}^{*}}
\newcommand{\bsd}{{A}_{s}^{*}}

\newcommand{\mps}{|\bar{A}_{p}|^2}

\newcommand{\bps}{|{A}_{p}|^2}

\newcommand{\phimn}{\exp{(-i \Delta k z)}}

\newcommand{\nth}{n_{\rm{th}}}

\newcommand{\gz}{\gamma_0}
\newcommand{\gp}{\gamma_{\Omega}}
\newcommand{\gn}{\gamma_{-\Omega}}
\newcommand{\gpp}{\gamma_{2 \Omega}}
\newcommand{\gnn}{\gamma_{-2 \Omega}}
\newcommand{\srt}{\sqrt{2 {\rm Im} \{ \gp} \} }
\newcommand{\srn}{\sqrt{2 {\rm Im} \{ \gamma_{ \Omega \rightarrow 0} \}  }}

\newcommand{\nn}{\nonumber}
\begin{document}

\title{Raman-noise induced quantum limits for $\chi^{(3)}$ nondegenerate phase-sensitive amplification
and quadrature squeezing}

\author{Paul L. Voss, Kahraman G. K\"{o}pr\"{u}l\"{u}, and Prem Kumar}

\address{Center for Photonic Communication and Computing, \\
ECE Department, Northwestern University,  \\
2145 Sheridan Road, Evanston, IL 60208-3118}

\begin{abstract}
We present a quantum theory of nondegenerate phase-sensitive
parametric amplification in a $\chi^{(3)}$ nonlinear medium.   The
non-zero response time of the Kerr $(\chi^{(3)})$ nonlinearity
determines the quantum-limited noise figure of $\chi^{(3)}$
parametric amplification, as well as the limit on quadrature
squeezing. This non-zero response time of the nonlinearity
requires coupling of the parametric process to a
molecular-vibration phonon bath, causing the addition of excess
noise through spontaneous Raman scattering. We present analytical
expressions for the quantum-limited noise figure of frequency
non-degenerate and frequency degenerate $\chi^{(3)}$ parametric
amplifiers operated as phase-sensitive amplifiers. We also present
results for frequency non-degenerate quadrature squeezing. We show
that our non-degenerate squeezing theory agrees with the
degenerate squeezing theory of Boivin and Shapiro as degeneracy is
approached.  We have also included the effect of linear loss on
the phase-sensitive process.
\end{abstract}

\maketitle 


\section{Introduction}

Fiber-optical parametric amplifiers (FOPAs) are currently the
subject of much research for use in wavelength
conversion\cite{wong03} and efficient broadband
amplification\cite{hansryd02}.  They are also candidates for
performing all-optical network functions\cite{Su00,Wang01,Wang02}.
Advances in pumping techniques have permitted improvements of the
noise figure (NF) of FOPAs operated
phase-insensitively~\cite{wong03,blows} and the manufacture of
high-nonlinearity and microstructure fibers has improved the gain
slope\cite{devgan03,radic03} of FOPAs.

In order to explain our experimental noise figure result for a {
FOPA operated as a phase-insensitive amplifier} (PIA)~\cite{voss03},
we have recently published a quantum theory of $\chi^{(3)}$
parametric amplifiers that takes into account the non-instantaneous
nonlinear response of the medium and the requisite addition of noise
caused by this non-instantaneous nonlinear
response~\cite{vosslimit,vosslong}. { The FOPA operates as a PIA
when an idler beam is generated wholly in the fiber or when the
phase difference (twice the phase of the pump less the phases of the
Stokes and anti-Stokes beams) is approximately $0$ or $\pi$ and the
Stokes and anti-Stokes beams are approximately equal in amplitude.
Our work with FOPAs} operated as PIAs also provides analytical
expressions for the noise figure of $\chi^{(3)}$ phase-insensitive
parametric amplifiers~\cite{vosslimit} and wavelength
converters~\cite{vosslong}.  This theory shows excellent agreement
with experiment~\cite{voss03}.  In addition, we have recently
experimentally investigated the noise figure spectrum for PIA and
wavelength converter operation of a FOPA, and shown good agreement
to an extended theory that includes distributed loss\cite{tang04}. {
This inclusion of distributed loss was necessary to model the
experiment\cite{tang04}, but also provides the necessary quantum
theory for predicting the performance of a distributed amplifier.}

Phase-sensitive amplifiers
(PSA)~\cite{PSAdegenrate1991,PSAdegenerate1994} are also of interest
because unlike PIAs, they can ideally provide amplification without
degrading the signal-to-noise ratio (SNR) at the
input\cite{caves82}. { Operation of a FOPA as a PSA occurs when the
phase difference at the input is approximately $\pi/2$ and the
signal and idler are approximately equal in amplitude.} Experiments
with fully frequency degenerate fiber phase-sensitive amplifiers
have demonstrated a noise figure of $2.0$ dB at a gain of $16$
dB~\cite{imajuku00}, a value lower than the standard
phase-insensitive high-gain $3$-dB quantum limit. A noise figure
below the standard PIA limit has also been reached in a low-gain
phase-sensitive amplifier~\cite{levandovsky01}. However, these fully
frequency degenerate PSA experiments were impaired by
guided-acoustic-wave Brillouin scattering (GAWBS)\cite{shelby85}
requiring pulsed operation~\cite{bergman94} or sophisticated
techniques for partially suppressing GAWBS~\cite{levandovsky01}. In
order to avoid the GAWBS noise one may obtain phase-sensitive
amplification with an improved experimental noise figure by use of a
frequency nondegenerate PSA. In addition, the nondegenerate PSA,
unlike its degenerate counterpart,  can be used with multiple
channels of data. A nondegenerate PSA is realized by placing the
signal in two distinct frequency bands symmetrically around the pump
frequency with a separation of several GHz, so that GAWBS noise
scattered from the pump is not in the frequency bands of the signal.
Such frequency nondegenerate regime has been demonstrated
 experimentally{, showing good agreement with theory for the average values of the signal and idler\cite{renyong_psa}.
 We have also experimentally demonstrated an improved bit-error rate
 by use of a PSA as opposed to a PIA of comparable gain\cite{renyong_el}.
 However, quantum-limited noise figure measurements have yet to be performed}. So an analysis of this case is
practically useful. Accordingly, we here describe in suitable detail
a quantum theory of FOPAs that takes into account the non-zero
response time of the $\ct$ nonlinearity along with the effect of
distributed linear loss. We present analytical expressions for the
quantum-limited noise figure of CW $\chi^{(3)}$ PSAs in the
frequency nondegenerate case. We also report the limiting value of
the NF when degeneracy is approached.  { In addition, we note that
this theory is valid for FOPAs used as distributed phase-sensitive
amplifiers \cite{vasilyev05}.}

A frequency nondegenerate parametric amplifier can also operate as a
phase-sensitive deamplifier (PSD) of  two-frequency input signals. {
This occurs when the phase-difference at the input is
 approximately $-\pi/2$ and the Stokes and anti-Stokes inputs are approximately
equal in amplitude.} When a PSD is operated with no input signal,
such a parametric amplifier is said to produce ``quadrature-squeezed
vacuum" (parametric fluorescence of the PSD) whose two-frequency
homodyne detection exhibits photocurrent variance less than that of
the vacuum for suitable choice of homodyne phases~\cite{yuen76}.
Quadrature squeezing has been proposed for applications in quantum
communications~\cite{yuen78,yuen79,yuen80}, improved measurement
sensitivity~\cite{caves81,treps03}, and { quantum
lithography\cite{boto2000}}.
 In the case of FOPAs, previous work by Shapiro and Boivin~\cite{Shapiro95} used the dispersionless theory
of self-phase modulation developed by Boivin and K\"{a}rtner
~\cite{boivin,kartner94} that included the non-instantaneous
response of the $\chi^{(3)}$ nonlinearity to obtain a limit on
quadrature squeezing in the fully four-degenerate-wave case. In this
paper, we present results for frequency nondegenerate CW quadrature
squeezing for a noninstantaneous nonlinearity in the presence of
dispersion. We show that optimal squeezing occurs for slightly
different input conditions than those for optimal classical
deamplification. In addition, we show that, unlike the
dispersionless case, the degree of squeezing reaches a constant
value in the long-interaction-length limit when the linear
phase-mismatch is nonzero and a non-instantaneous nonlinear response
is present. Our nondegenerate squeezing theory agrees with the
previous degenerate squeezing results of Boivin and
Shapiro~\cite{Shapiro95} when degeneracy is approached.

This paper is organized as follows:  In Section 2 we discuss the
solution of the equations describing evolution of the mean values of
the pump, Stokes, and anti-Stokes fields.  In Section 3, we present
a quantum mechanically consistent theory of the FOPAs. { For
calculation of the noise figure in phase-sensitive operation, we
need to obtain only the mean and variance of the total photocurrent
at the Stokes and anti-Stokes wavelengths. Thus we chose to
calculate only the output Heisenberg annihilation operators for the
Stokes and anti-Stokes frequency-pair of interest.  We thus obtain
the desired results in a simpler way than had we used the Wigner or
positive-P formalism developed by Drummond and
Corney\cite{drummond01} for propagation of the quantum states of
pulses in the fiber.} In Sections 4 and 5, we apply this theory to
obtain the noise figure of phase-sensitive amplification and to
obtain the degree of nondegenerate quadrature squeezing,
respectively. We reemphasize the main results and conclude in
Section 6.

\section{Classical phase-sensitive amplification and deamplification}

 {We have discussed the $\chi^{(3)}$ nonlinear response at length~\cite{vosslong}; only a very brief summary
 is presented here.

 The nonlinear refractive index of the Kerr interaction can be
written as \begin{equation} n_2=\frac{3 \chi^{(3)}}{4\epsilon_0
n_0^2 c},\end{equation} where} $n_0$ is the linear refractive
index of the nonlinear material, $\epsilon_0$ is the permittivity
of free space, and $c$ is the speed of light in free space. For
clarity, we state that
\begin{eqnarray} \ct \equiv \ct_{1111} \left[\frac{{\rm m}^2}{{\rm
V}^2} \right].\end{eqnarray} The $\ct(t)$ nonlinear response is
composed of a time-domain delta-function-like electronic response
{($\ll 1 \, {\rm fs}$)} that is constant in the frequency domain
over the bandwidths of interest and a time-delayed Raman response
{($\approx 50 \, {\rm fs}$)} that varies over frequencies of
interest and is caused by back action of nonlinear nuclear
vibrations on electronic vibrations. Recent experimental and
theoretical results demonstrate that the nonlinear response function
$\ct(t)$ can be treated as if it were real in the time
domain,{~\cite{newbury,newbury03}} yielding a real part that is
symmetric in the frequency domain with respect to pump detuning and
an imaginary part in the frequency domain that is anti-symmetric.

Although a nonlinear response is also present in the polarization
orthogonal to that of the pump, this cross-polarized nonlinear
interaction is ignored because we assume that the pump, Stokes, and
anti-Stokes fields of interest stay copolarized { as their
polarization states evolve during propagation through the FOPA.}
Parametric fluorescence and Raman spontaneous emission are present
in small amounts in the polarization perpendicular to the pump, but
do not affect the NF of the amplifier.

We can write $N_2(\Omega)$ in the frequency domain as a sum of
electronic and molecular contributions:
\begin{equation}
N_2(\Omega)= n_{2 \rm{e}} + n_{2 \rm{r}}F(\Omega). \label{fth}
\end{equation}
We also define a nonlinear coefficient $\gp$ to be
 \begin{equation} { \gp=\frac{2 \pi
N_2(\Omega)}{ \lambda A_{\rm{eff}}} { \left[ \frac{1}{W \cdot m}
\right]}}, \label{hwdef}
\end{equation} where $\lambda$ is the pump wavelength and
$A_{\rm{eff}}$ is the fiber effective area. {Thus our $\gz$ is
equivalent to the nonlinear coupling coefficient $\gamma$ used in
Agrawal \cite{agrawal}.} It is the scaling of $A_{\rm eff}$ with
wavelength that mainly causes $\gp$ to be no longer anti-symmetric
with $\gn$ at detunings greater than several THz.  In what follows,
our analytical treatment of the mean fields allows for the more
general case of asymmetry in the Raman-gain spectrum. However, other
results including graphs assume an anti-symmetric Raman spectrum as
this has a minor effect on the quantum noise at large detunings.

We next present solutions to the mean field equations  governing a
parametric amplifier. The optical fields are assumed to propagate in
a dispersive, polarization-preserving, single-transverse-mode fiber
under the slowly-varying-envelope approximation.  As the involved
waves are quite similar in frequency, to good approximation all
fields can be treated as if their transverse mode profiles are
idententical.  Even though the fibers used to construct FOPAs
typically support two polarization modes and the polarization state
of the waves is usually  elliptical at a given point $z$ in the
FOPA, for typical fibers it is still appropriate to describe the
system with a scalar theory if the detuning is relatively
small\cite{tang04}. This is because the input waves are copolarized
at the beginning of the amplifier and the fields of interest remain
essentially copolarized during propagation down the fiber.

Consider the field
\begin{eqnarray}
{A}(t)={A}_{p} + {A}_{s}\exp(i \Omega t) + {A}_{a}\exp(-i \Omega t)
\label{eq3freq} \end{eqnarray} for the total field propagating
through a FOPA having a frequency and polarization degenerate pump.
{ Here $|A|^2$ has units of Watts.}The lower frequency field we
refer to as the Stokes field, ${A}_{s}$; the higher frequency field
referred to as the anti-Stokes field, ${A}_{a}$. The classical
equation of motion for the total field\cite{golovchenko90} with the
addition of arbitrary frequency dependent loss is:
\begin{equation}
\frac{\partial {A}(t)}{ \partial z} =i \left[ \int {\textrm{ d}}\tau
\, \gamma(t-\tau){A}^{*}(\tau){A}(\tau) \right] {A}(t) - \int
\frac{\alpha(\Omega)} {2} \widetilde{A}(\Omega)\exp(-i \Omega t) \,
{\rm d}\Omega, \label{classicalfullwave}
\end{equation}
where $\alpha(\Omega)$ is the power attenuation coefficient at
detuning $\Omega$ from the pump and $\widetilde{A}(\Omega)$ is the
Fourier transform of the field. Because the involved waves
(Stokes, anti-Stokes, and pump) are CW, the usual group-velocity
dispersion term does not explicitly appear in
Eq.~(\ref{classicalfullwave}). However, dispersion is included;
its effect is simply to modify the wavevector of each CW
component. Taking the Fourier transform of
Eq.~(\ref{classicalfullwave}) and separating into
frequency-shifted components that are capable of phase-matching,
we obtain the following differential equations for the mean
fields~\cite{golovchenko90}:
\begin{eqnarray}
\frac{{\rm{d}} \bpn }{{\rm{ d}}z} &=& i \, \gz \, \bps \bpn -
\frac{\alpha_p}{2}{A}_p, \label{meanfieldp} \\ \frac{{\rm{d}} \ban
}{{\rm{d}}z} &=& i \, ( \gz + \gp ) \, \bps \ban + i \gp \bpn^2 \bsd
\phimn  -
\frac{\alpha_a}{2}\ban \label{meanfielda},\\
\frac{{\rm{d}} \bsn }{{\rm{d}}z} &=& i \, ( \gz + \gn ) \, \bps \bsn
+ i \gn \bpn^2 \bad \phimn  - \frac{\alpha_s}{2}\bsn.
\label{meanfields}
\end{eqnarray}
{ Here  $\Delta k= k_a + k_s - 2 k_p$ is the phase mismatch.
Expanding the wavevectors in a Taylor series around the pump
frequency to second order, one obtains $\Delta k = \beta_2
\Omega^2$ to second order, where $\beta_2$ is the group-velocity
dispersion coefficient.  The attenuation coefficients are
$\alpha_j$ for $j= p,a,s$ at the pump, anti-Stokes, and Stokes
wavelenths, respectively.  The nonlinear coupling coefficients
$\gz$, $\gp$, and $\gn$ are as defined in the previous section.
 Eqs.~(\ref{meanfieldp}--\ref{meanfields})} are valid when the pump
remains essentially undepleted by the Stokes and anti-Stokes waves
and is much stronger than the Stokes and anti-Stokes waves. The
solution to Eqs.~(\ref{meanfieldp}) and (\ref{meanfields}) can be
expressed as
\begin{eqnarray}
{A}_a(z,L)&=& \mu_a(z,L) {A}_a(z) + \nu_a(z,L) {A}_s^*(z), \\
{A}_s(z,L)&=& \mu_s(z,L) {A}_s(z) + \nu_s(z,L) {A}_a^*(z),
\end{eqnarray}
where we have explicitly written the solution as a function of
both a starting point $z$ for the parametric process and an end
point $L$ for the fiber. We do this because we will be interested
not only in the input-output relationships of the electromagnetic
fields, i.e. $A_a(0,L)$, but also evolution of noise generated at
a point $z$ that propagates to the end of the fiber, $L$. In the
following subsections, we provide expressions for $\mu_j(z,L)$ and
$\nu_j(z,L)$ for the three main cases of interest.

\subsection{Distributed loss solution}
In the most general case, when there are no restrictions on
$\Delta k$ and distributed linear loss is present,
Eqs.~(\ref{meanfieldp}--\ref{meanfields}) can be shown to have a
series solution.  We here briefly outline the derivation of this
solution. Solving for the mean field of the pump,
Eq.~(\ref{meanfieldp}), we obtain
\begin{eqnarray}
\bpn(z)=\exp[i \gz I_{p}(0) z_{\rm eff} - \frac{\alpha_{p} z}{2} ]{
\bpn(0)},
\end{eqnarray}
where the effective length $z_{\rm eff}$ is defined to be $z_{\rm
eff}=[1-\exp(-\alpha_p z)]/ \alpha_p$. Further defining the initial
pump power in Watts to be $I_p(0)=|\bpn (0)|^2$, and setting the
reference phase to be that of the pump at the input of the fiber, we
substitute the  resulting expressions into Eqs.~(\ref{meanfielda})
and (\ref{meanfields}). Writing
\begin{eqnarray}
\Man &=& {B}_a \exp [ i (\gz + \gp) I_p(0) z_{\rm eff}
- \alpha_a z /2 ], \\
\Msn &=& {B}_s \exp [ i (\gz + \gn) I_p(0) z_{\rm eff} - \alpha_s z
/2 ],
\end{eqnarray}
and making a change of variable from $z$ to $z_{\rm eff}$, one
obtains
\begin{eqnarray}
\frac{{\rm{d}} {B}_a }{{\rm{d}}z_{\rm eff}} &=& i \gp I_p(0)
\exp[-f(z_{\rm eff})]
{B}_s^* \label{meanfieldab},\\
\frac{{\rm{d}} {B}_s^* }{{\rm{d}}z_{\rm eff}} &=& -i \gn^* I_p(0)
\exp[f(z_{\rm eff})] {B}_a \label{meanfieldsb},
\end{eqnarray}
where
\begin{eqnarray} f(z_{\rm eff})=i[\gp + \gn^*] I_p(0)
z_{\rm eff} - \frac{(\alpha_s - \alpha_a + 2 i \Delta
k)\ln(1-\alpha_p z_{\rm eff}) }{2 \alpha_p}. \end{eqnarray} After
some algebra and making use of the substitutions
\begin{eqnarray}
\bar{F}_a &=& {B_a} \exp[f(z_{\rm eff})], \\
\bar{F}_s^* &=& {B}_s^* \exp[-f(z_{\rm eff})],
\end{eqnarray}
we can obtain the nonlinear coupled equations
\begin{eqnarray}
\frac{{\rm{d}} {F}_a }{{\rm{d}}z_{\rm eff}} - [\Gamma + \frac{\Lambda}{1-\alpha_p z_{\rm eff}}]{F}_a &=& \xi_1 {F}_s^* \label{meanfieldaf},\\
\frac{{\rm{d}} {F}_s^* }{{\rm{d}}z_{\rm eff}} +[\Gamma +
\frac{\Lambda}{1-\alpha_p z_{\rm eff}}]\bar{F}_s^* &=& \xi_2 {F}_a,
\label{meanfieldsf} \end{eqnarray}  where the following constants
are used for calculating evolution from point $z$ to $L$:
$\Gamma=i[\gp+\gn^*]I_p(z)/2$, $\Lambda=[\alpha_s/2 -\alpha_a/2 + i
\Delta k]/2$, $\xi_1=i \gp I_p(z)$, and $\xi_2=-i \gn^* I_p(z)$.
Using the expansion
\begin{eqnarray} \frac{1}{1- \alpha_p z_{\rm
eff}}=\sum_{n=0}^{\infty} \alpha_p^n z_{\rm eff}^n,
\label{expansion}
\end{eqnarray} on the nonlinear term in Eqs.~(\ref{meanfieldaf}) and (\ref{meanfieldsf}), we find a series
solution for ${F}_j$ and then obtain ${A}_j$.

The series solution converges in relatively few terms when
$\alpha_p z_{\rm eff}$ is small, which is the case for practical
amplifiers. The  following solutions are then obtained:
\begin{eqnarray} \mu_a(z,L) &=& \exp[p(z,L)] \sum_{n=0}^{\infty} a_n L_{\rm eff}^n \,\,\,\,\,({a_0=1,s^*_0=0}), \label{mu_a} \\
\mu_s(z,L) &=& \exp[p(z,L)] \sum_{n=0}^{\infty}s_n L_{\rm eff}^n \
\,\,\,\,\,{(a_0=0,s^*_0=1)}, \label{mu_s} \\ \nu_a(z,L) &=&
\exp[p(z,L)]\sum_{n=0}^{\infty}a_n L_{\rm eff}^n
\,\,\,\,\,({a_0=0,s^*_0=1}),
 \label{nu_a} \\
  \nu_s(z,L) &=&  \exp[ p(z,L)] \sum_{n=0}^{\infty}s_n L_{\rm eff}^n
\,\,\,\,\,({a_0=1,s^*_0=0}), \label{nu_s}\end{eqnarray} where
\begin{eqnarray}  p(z,L) &=& i[\gz+ (\gp - \gn^*)/2]I_p(z) L_{\rm eff}
\\ & & -i \Delta k (L-z)/2 -\alpha_a (L-z)/4 -\alpha_s (L-z)/4\end{eqnarray}
and $L_{\rm eff}= \{ 1-\exp[-\alpha_p(L-z)] \} / \alpha_p$. The
coefficients $a_n$ and $s^*_n$ are then calculated through the
following recursion relations:
\begin{eqnarray}
a_n &=& \frac{\Gamma a_{n-1}+ \xi_1 s^*_{n-1} + \Lambda
\sum_{j=0}^{n-1} \alpha_p^j a_{n-1-j} }{n}, \\
s^*_n &=& \frac{-\Gamma s^*_{n-1} + \xi_2 a_{n-1} - \Lambda
\sum_{j=0}^{n-1} \alpha_p^j s^*_{n-1-j}  }{n}.
\end{eqnarray}

\subsection{Lossless, $\Delta k \neq 0$ solution}

The solution for lossless fiber and $\Delta k \neq 0$ is well
known, as are the $\mu$ and $\nu$ functions which can be expressed
as\cite{golovchenko90}
\begin{eqnarray} \mu_a(z,L)&=& \exp \left( - \frac{i \left(\Delta k
- [2 \gz + \gp - \gn^*] I_p \right)(L-z) }{2} \right) \nonumber \\
& & \times \left( \frac{ i \kappa}{2g} \sinh[g(L-z)] +
\cosh[g(L-z)] \right), \label{mu_a}
\end{eqnarray}\begin{eqnarray} \mu_s(z,L) &=& \exp\left(- \frac{i
\left(\Delta k - [2 \gz + \gn - \gp^*] I_p \right)(L-z)
}{2}\right) \nonumber \\
& &  \left( \frac{ i \kappa^*}{2g^*} \sinh[g^*(L-z)] +
\cosh[g^*(L-z)]\right), \label{mu_s}\end{eqnarray}\begin{eqnarray}
\nu_a(z,L)&=& \exp \left(- \frac{i \left(\Delta k - [2 \gz + \gp -
\gn^*]  I_p \right)(L-z) }{2} \right) \frac{i \gp { A_p(z)^2}{}}{g}
\sinh[ g (L-z)], \nonumber \\ \label{nu_a}
\end{eqnarray}
\begin{eqnarray} \nu_s(z,L) &=& \exp \left( - \frac{i \left(\Delta k - [2
\gz + \gn - \gp^*] I_p \right)(L-z) }{2} \right) \frac{i \gn {
A_p(z)^2}{}}{g^*} \sinh[ g^* (L-z)]. \nonumber \\ \label{nu_s}
\end{eqnarray} Here  $I_p(z)=I_p=|{A}_p(0)|^2$ is the pump power in
{ Watts}, $\kappa = \Delta k + ( \gp + \gn^*)I_p$, and $g= \sqrt{
- (\kappa/2)^2 + \gp \gn^* I_p^2}$ is the complex gain
coefficient.

\subsection{Lossless, $\Delta k =0$ solution}
We also state the results for the lossless, $\Delta k=0$ case,
which is useful in our analysis near degeneracy. We have $\Delta
k=0$ when the FOPA is pumped at the zero dispersion wavelength or
if the system is treated as if dispersionless.  When the three
frequencies are very nearly degenerate, we can also make the
approximation that $\Delta k=0$. Then the $\mu$ and $\nu$
functions become
\begin{eqnarray}
\mu_a(z,L) &=& \exp \left[ i \left( \frac{2 \gz + \gp -\gn^*}{2}
\right) I_p (L-z) \right] [1 + i \gp { (L-z)} I_p],
\label{mua_dkz} \\
\mu_s(z,L) &=& \exp \left[ i \left( \frac{2 \gz + \gn -\gp^*}{2}
\right) I_p (L-z) \right] [1 + i \gn {(L-z)} I_p],
 \label{mus_dkz}\\
\nu_a(z,L) &=& \exp \left[i \left( \frac{ 2 \gz + \gp -\gn^*}{2}
\right) I_p (L-z) \right] i \gp {(L-z) A_p(z)^2},
 \label{nua_dkz}\\
\nu_s(z,L) &=& \exp \left[ i \left( \frac{2 \gz + \gn -\gp^*}{2}
\right) I_p (L-z) \right] i \gn {(L-z)A_p(z)^2}. \label{nus_dkz}
\end{eqnarray}
Under the normal assumption of an anti-symmetric Raman gain profile
($\gp=\gn^*$), we see the power gain { for an input only on the
anti-Stokes side}, $G_{anti-Stokes}=|\mu_a(0,L)|^2 = 1 - 2 {\rm
Im}\{ \gp \} I_p L + |\gp|^2 I_p^2 L^2$, has a four wave mixing gain
that is quadratic as a function of fiber length and that the Raman
loss at the anti-Stokes wavelength is linear in pump power and
length ($-2 {\rm Im}\{ \gp \} I_p L$). Similarly, the Raman gain at
the Stokes wavelength is linear in power and length ($2 {\rm Im}\{
\gn \} I_p L$).

\subsection{Optimal classical phase-sensitive amplification and deamplification}
We next find the optimal  phase-sensitive amplification and
phase-sensitive deamplification of a mean field consisting of a
superposition of Stokes and anti-Stokes fields.  We define optimal
phase-sensitive amplification (deamplification) as the greatest
(least) output signal power possible for a fixed amount of input
signal power. Assuming coherent signal inputs $\zeta_j$ having
powers $|\zeta_j|^2$ and phases $\exp(i \theta_j)$ for $j=a,s$,
the phase-sensitive gain of the PSA is
\begin{eqnarray}
G &=& \frac{|{A}_a(L)|^2 + |{A}_s(L)|^2}{|{A}_a(0)|^2 +
|{A}_s(0)|^2} \nonumber
\\ &=&
\frac{|\mu_a \zeta_a + \nu_a \zeta_s^*|^2 + |\mu_s \zeta_s + \nu_s \zeta_a^*|^2}{|\zeta_a|^2 +|\zeta_s|^2}\nonumber \\
&=&  \frac{ (|\mu_a|^2+ |\nu_s|^2)|\zeta_a|^2 + ( |\nu_a|^2 +
|\mu_s|^2) |\zeta_s|^2   + \left\{ [(\mu_s \nu_s^* + \mu_a
\nu_a^*) |\zeta_a||\zeta_s| \,\, e^{i(\theta_a+\theta_s)}] + {\rm
c.c.} \right \} }{|\zeta_a|^2+|\zeta_s|^2}. \nonumber \\
\label{psa_mean_out}
\end{eqnarray}
By properly choosing the relative power of the Stokes and
anti-Stokes inputs and their sum phase $\theta = \theta_a +
\theta_s$ relative to the input pump phase,  one achieves
maximimum (minimimum) phase-sensitive amplification
(deamplification). The optimum sum phases $\theta_{\rm psa,
\,\,opt}$ and $\theta_{\rm psd, \,\,opt}$ are
\begin{eqnarray} \theta_{\rm psa , \,\,opt}=-\arg[\mu_s \nu_s^* + \mu_a \nu_a^*], \label{theta_psa} \\
                 \theta_{\rm psd , \,\,opt}=\pi -\arg[\mu_s \nu_s^* + \mu_a
                 \nu_a^*],
         \end{eqnarray}
for amplification and deamplification, respectively. By setting
the sum input power $|\zeta_a|^2 + |\zeta_s|^2 = \mathcal{C}$ to
be some constant $\mathcal{C}$, the extrema of
Eq.~(\ref{psa_mean_out}) can be found to occur when the proportion
of input anti-Stokes power to the total input power is
\begin{eqnarray}
  \frac{|\zeta_a|^2}{|\zeta_a|^2 + |\zeta_s|^2}=\frac{1}{2}\left ( 1 \pm \frac{|\mu_s|^2 - |\mu_a|^2}
  {\sqrt{4|\mu_a \nu_a^* + \mu_s \nu_s^*|^2 +
  [|\mu_s|^2-|\mu_a|^2]^2}}
\right) \label{class_split} \end{eqnarray} where the negative root
corresponds to optimum phase-sensitive amplification, the positive
root to optimum phase-sensitive deamplification.  The maximum PSA
gain, $G_{\rm PSA}$ is found by insertion of
Eqs.~(\ref{theta_psa}) and (\ref{class_split}) into
Eq.~(\ref{psa_mean_out}).  The result simplifies to:
\begin{eqnarray}
G_{\rm PSA} &=&  \frac{( |\mu_a|^2 +  |\mu_s|^2 +  |\nu_a|^2 +
|\nu_s|^2)}{2} + \frac{\sqrt{4|\mu_a \nu_a^* + \mu_s \nu_s^*|^2 +
(|\mu_s|^2 - |\mu_a|^2)^2}}{2} \nonumber \\ & &+ \frac{(|\mu_s|^2
- |\mu_a|^2)(|\nu_a|^2 - |\nu_s|^2)}{\sqrt{4|\mu_a \nu_a^* + \mu_s
\nu_s^*|^2 + (|\mu_s|^2 - |\mu_a|^2)^2}}.
\end{eqnarray}

In Fig.~\ref{psa_mean2} all of the plots are optimal in the sense
that the best total phase and relative power of the input fields is
chosen. It is clear that the Raman effect is negligible as those
curves including the Raman effect (circles and squares) are very
similar to those neglecting it (solid curve and dashed curve). This
plot shows also that the definition of the effective length is a
mathematical one and not a good guide for estimating the gain
profile. The dotted and dash-dot curves show the gain spectrum of a
lossy fiber ($0.41$ dB$/$km) $4.44$ km in length. The other curves
are for a lossless fiber of effective length $L_{\rm eff}=3.63$ km.
Thus distributed loss has a greater effect than might be supposed:
fibers experience noticeably less gain than lossless fibers do when
the lossless fiber has a length equal to the effective length of a
lossy fiber.

Some of the characteristics of the classical phase-sensitive
response can be seen in Fig.~\ref{psa_mean1}, which is a plot of the
phase-sensitive gain vs. fiber length. The primary feature of
phase-sensitive amplification is that the mean-field gain is
relatively insensitive to the relative strength of the two input
fields (this can be seen by the overlap of the thin solid lines with
the squares and circles).  In addition, typical distributed losses
do not significantly impact the gain of the fiber, as can be seen by
comparison of the squares (lossless fiber) with the circles, which
represent fiber with loss of $0.25$ dB$/$km.  On the other hand, the
achievable degree of phase-sensitive deamplification is much more
sensitive to the relative proportion of the input fields which can
be seen by comparison of the dash-dotted and dashed lines (equal
power splitting, i.e., $|\zeta_a|^2=|\zeta_s|^2$) with the dotted
and thick solid line (optimum relative proportion). In addition,
distributed losses also set a limit on classical deamplification as
can be seen by comparison of the dashed line (lossy) with the
dash-dotted lines (lossless) and comparison of the dotted line
(lossy) with the thick solid line (lossless).

\section{Input-output quantum mode transformations}

In this section, we discuss the quantum mechanics of the
$\chi^{(3)}$ parametric amplifier { with a strong, undepleted,
coherent-state pump} and derive input-output mode transformations in
the Heisenberg picture which can be used to calculate the noise
figure of the phase-sensitive operation of a FOPA and the
accompanying quadrature squeezing. Here we also extend our
previously described quantum theory\cite{vosslimit,vosslong} to
include the effects of loss.  {  In order to make our treatment
consistent with the customary formalism in quantum optics, we
rescale our field to a photon-flux field, i.e. the expectation of
the field  $\langle \hat{A} \rangle=\bar{A}=A/\sqrt{\hbar \omega}$
so that $|\bar{A}|^2$  has units of photons/sec.  We also rescale
our nonlinear coefficient $\bar{\gamma}_{\Omega} = \gp \hbar
\omega$, where $\hbar$ is Planck's constant over $2 \pi$.}

\renewcommand{\gz}{\bar{\gamma}_{0}}
\renewcommand{\gp}{\bar{\gamma}_{\Omega}}
\renewcommand{\gn}{\bar{\gamma}_{-\Omega}}
\renewcommand{\gpp}{\bar{\gamma}_{2 \Omega}}
\renewcommand{\gnn}{\bar{\gamma}_{-2 \Omega}}
\renewcommand{\srt}{\sqrt{2 {\rm Im} \{ \gp} \} }
\renewcommand{\srn}{\sqrt{2 {\rm Im} \{ \gamma_{ \Omega \rightarrow 0} \}  }}

{ We begin with the} { continuous-time} quantum equation of motion
for a multi-mode field in the presence of a non-instantaneous
nonlinearity.   This model was presented by Carter and
Drummond~\cite{drummond01,Carter87}, solved for the case of
dispersionless self-phase modulation by Boivin~\cite{boivin}, and
also derived in detail by K\"{a}rtner~\cite{kartner94}.  We have
\begin{eqnarray}
\frac{\partial \hat{A}(t)}{ \partial z} &=& - \int
\frac{\alpha(\Omega)} {2} \widetilde{\hat{A}}(\Omega)\exp(-i \Omega
t) \, {\rm d}\Omega + i \left[ \int {\textrm{ d}}\tau \, {
\bar{\gamma}}(t-\tau)\hat{A}^{\dagger}(\tau)\hat{A}(\tau) \right]
\hat{A}(t) \nonumber \\ &+& i \, \hat{m}(z,t)\hat{A}(t) { +
\hat{l}(z,t)} \label{fullwave},
\end{eqnarray}

{ wherein ${ \bar{\gamma}(t)}$ is the causal response function of
the nonlinearity, i.e., the inverse Fourier transform of ${ \gp}$ in
Eq.~(\ref{hwdef}).  In Eq.~(\ref{fullwave}), $\hat{m}(z,t)$ is a
Hermitian phase-noise operator

\begin{eqnarray}{ \hat{m}(z,t) = \int_0^{\infty} {\rm{d}}\Omega
\,\, \frac{\sqrt{2 {\rm Im} \{ \gp \} }}{{2 \pi}} \{i \,
\hat{d}^{\dagger}_{\Omega}(z)e^{i \, \Omega t} - i\,
\hat{d}_{\Omega}(z)e^{-i \Omega t}  \}, \label{eq:pno}}
\end{eqnarray}
 which describes coupling of the
field to a collection of localized, independent, medium oscillators
(optical phonon modes).  { In addition, $\hat{l}(t)$} is a noise
operator
\begin{eqnarray}
{ \hat{l}(t)=\int_{- \infty}^{\infty} {\rm d}\Omega
\frac{\sqrt{\alpha_{\Omega}}}{2 \pi}
 \exp(i \Omega t) \hat{v}_{\Omega}(t)},
 \end{eqnarray}
which describes the coupling of the field to a collection of
localized, independent, oscillators in vacuum state. This coupling
is required to preserve the continuous-time commutators}
\begin{equation} [ \hat{A}(t) , \hat{A}^{\dagger}(t')] = \delta(t-t'), \end{equation} \begin{equation} [ \hat{A}(t) ,
\hat{A}(t') ] = 0.
\end{equation}
 Note that the time $t$ is in a reference frame traveling at
group velocity $v_g$, i.e., $t = t_{\rm{stationary\,\,frame}} -
\frac{z}{v_g}$. The weighting function ${ \sqrt{2 {\rm Im} \{ \gp \}
}}$ must be positive for $\Omega>0$ so that the molecular vibration
oscillators absorb energy from the mean fields rather than providing
energy to the mean fields.  The operators $\hat{d}_{\Omega}(z)$ and
$\hat{d}^{\dagger}_{\Omega}(z)$ obey the commutation relation
\begin{eqnarray}
[\hat{d}_{\Omega}(z),\hat{d}^{\dagger}_{\Omega'}(z')]=\delta(\Omega-\Omega')\delta(z-z')
\label{eq:thcommrel}
\end{eqnarray}
and each phonon mode is in thermal equilibrium:
\begin{eqnarray}
\langle \hat{d}^{\dagger}_{\Omega}(z)\hat{d}_{\Omega'}(z')\rangle=
\delta(\Omega-\Omega') \delta(z-z')n_{\rm{th}} \label{eq:thmean}
\end{eqnarray}
with a mean phonon number  $n_{\rm{th}}=[\exp(\hbar
\Omega/kT)-1]^{-1}$.  Here $\hbar$ is Planck's constant over $2
\pi$, $k$ is Boltzmann's constant, and $T$ is the temperature. The
operators corresponding to vacuum modes mixing into the Stokes and
anti-Stokes frequencies, $\hat{v}_{\pm \Omega}$, obey the
commutation relations
\begin{eqnarray}
[\hat{v}_{\pm \Omega}(z),\hat{v}^{\dagger}_{\pm
\Omega}(z)]=\delta(\pm \Omega- \pm \Omega')\delta(z-z')
\end{eqnarray}
and have no photons in them, i.e., $\langle \hat{v}^{\dagger}_{\pm
\Omega}(z)\hat{v}_{\pm \Omega'}(z')\rangle=0$.

{ We assume that the total field present at the input of the fiber
contains only a single-frequency pump, a Stokes field, and a
symmetrically placed anti-Stokes field.  The operators corresponding
to these modes are
\begin{eqnarray} \hat{A}(t)=\hat{A}_{p} + \hat{A}_{s}\exp(i \Omega
t) + \hat{A}_{a}\exp(-i \Omega t). \label{eq3freq}
\end{eqnarray} We do not need to consider other modes in the fiber when the pump is strong ($|\bar{A}_p|^2 \gg |\bar{A}_a|^2 ,
|\bar{A}_s|^2)$ and remains undepleted and the Stokes and
anti-Stokes frequencies of interest are too weak to serve as pumps
to other nonlinear processes. When this latter condition occurs,
fluctuations from other higher order mixing modes are not coupled
in. In addition, the vacuum modes at the input do not grow due to
parametric fluorescence to become sufficiently strong to serve as
pumps for nonlinear processes. This is because the input Stokes and
anti-Stokes fields would saturate the pump long before this would
happen. Thus the symmetric pairing of the Stokes and anti-Stokes
fields of interest is justified and the pairs do not couple through
the nonlinear process to other frequencies.

In order to obtain the coupled-wave equations for the three
frequencies of interest, we insert Eq.~(\ref{eq3freq}) into
Eq.~(\ref{fullwave}), take the Fourier transform of the resulting
equation, and separate into different frequencies. The resulting
nonlinear quantum operator equations are:
\begin{eqnarray}
\frac{{\rm d}\opn}{{\rm d}z} &=& i \gz \opd \opn \opn + i[\gz +\gp]
\osd \osn \opn + i [ \gz + \gn ] \oad \oan \opn \nonumber \\ &+&
i[\gn + \gp] \opd \osn \oan \exp(i \Delta k z) -\srt \oan \hat{d}^{\dagger}_{\Omega}(z)\exp(-i (k_p - k_a) k z)  \nonumber \\
& & +\srt \osn \hat{d}_{\Omega}(z)\exp(-i (k_p -k_s) k z) + \srn
\opn
[\hat{d}_{\Omega \rightarrow 0+}(z)+\hat{d}^{\dagger}_{\Omega \rightarrow 0-}(z)] \nonumber \\
&-& \frac{\alpha_p}{2} \hat{A}_p + \sqrt{\alpha_p}\hat{v}_p(z), \label{fop}\\
\frac{{\rm d}\oan}{{\rm d}z} &=& i [\gz + \gp] \opd \opn \oan + i
\gz \oad \oan \oan + i[\gz + \gpp] \osd \osn \oan \nonumber \\ &+& i
\gp \opn^2 \osd \exp(-i \Delta k z) - \srt \opn \exp[i (k_p -k_a)z]
\hat{d}_{\Omega}(z)
\nonumber \\
&-& \frac{\alpha_a}{2} \oan + \sqrt{\alpha_a}\hat{v}_a(z), \label{foa} \\
\frac{{\rm d}\osn}{{\rm d}z} &=& i [\gz + \gn] \opd \opn \osn + i
\gz \osd \osn \osn + i[\gz + \gnn] \oad \oan \osn \nonumber \\ &+& i
\gn \opn^2 \oad \exp(-i \Delta k z) + \srt \opn \exp[i (k_p -k_s)z]
\hat{d}_{\Omega}^{\dagger}(z)
\nonumber \\
&-&  \frac{\alpha_s}{2} \osn + \sqrt{\alpha_s}\hat{v} _s(z).
\label{fos}
\end{eqnarray}
Eqs. (\ref{fop} - \ref{fos}) are nonlinear  in the quantum operators
(except for the last two terms of each equation) and are difficult
to solve. However, one may linearize these equations
around the mean values of the operators and obtain solutions
accurate to first order in the fluctuations. Using the definition
\begin{eqnarray}
\hat{A}_j= \bar{A}_j + \hat{a}_j, \label{lineq} \end{eqnarray} where
$\bar{A}_j$ represents a c-number mean value of $\hat{A}_j$ and
$\hat{a}_j$ represents the quantum fluctuations of $\hat{A}_j$, and
where $j=p,a,s$, we expand Eqs. (\ref{fop} - \ref{fos}) to obtain
the following equations for the mean values (i.e., those terms
containing no fluctuation operators):
\begin{eqnarray}
\frac{{\rm d}\Mpn}{{\rm d}z} &=& i \gz \Mpd \Mpn \Mpn + i[\gz
+\gp] \Msd \Msn \Mpn + i [ \gz + \gn ] \Mad \Man \Mpn \nonumber \\
& & +
i[\gn + \gp] \Mpd \Msn \Man \exp(i \Delta k z) - \frac{\alpha_p}{2} \Mpn, \label{fMp}\\
\frac{{\rm d}\Man}{{\rm d}z} &=& i [\gz + \gp] \Mpd \Mpn \Man + i
\gz \Mad \Man \Man + i[\gz + \gpp] \Msd \Msn \Man \nonumber \\ & & +
i
\gp \Mpn^2 \Msd \exp(-i \Delta k z) - \frac{\alpha_a}{2}, \Man\label{fMa} \\
\frac{{\rm d}\Msn}{{\rm d}z} &=& i [\gz + \gn] \Mpd \Mpn \Msn + i
\gz \Msd \Msn \Msn + i[\gz + \gnn] \Mad \Man \Msn \nonumber \\ & & +
i \gn \Mpn^2 \Mad \exp(-i \Delta k z) - \frac{\alpha_s}{2} \Msn.
\label{fMs}
\end{eqnarray}
The equations for those terms which contain one fluctuation operator
are 
\begin{eqnarray}
\frac{{\rm d}\fpn}{{\rm d}z} &=& i \gz[2|\Mpn|^2 \fpn + \Mpn^2 \fpd]
+ i[\gz +\gp] [|\Msn|^2 \fpn + \Msd \Mpn \fsn + \Msn \Mpn \fsd] \nonumber \\
& & + i [ \gz + \gn ] [|\Man|^2 \fpn + \Mad \Mpn \fan + \Man \Mpn \fad] \nonumber \\
& & +
i[\gn + \gp] [\Mpd \Msn \fan + \Mpd \Man \fsn + \Msn \Man \fpd] \exp(i \Delta k z) \nonumber  \\
& & -\srt \Man \hat{d}^{\dagger}_{\Omega}(z)\exp(-i (k_p - k_a) k z)
+ \srt \Msn \hat{d}_{\Omega}(z)\exp(-i (k_p -k_s) k z) \nn \\ & & +
\srn \Mpn
[\hat{d}_{\Omega \rightarrow 0+}(z)+\hat{d}^{\dagger}_{\Omega \rightarrow 0-}(z)] -\frac{\alpha_p}{2} \fpn + \sqrt{\alpha_p}\hat{v}_p(z), \label{ffp}\\
\frac{{\rm d}\fan}{{\rm d}z} &=& i [\gz + \gp][|\Mpn|^2
\fan + \Mpd \Man \fpn + \Mpn \Man \fpd] + i \gz [2 |\Man|^2 \fan + \Man^2 \fad] \nn \\
& & + i[\gz + \gpp][|\Msn|^2 \fan + \Msd \Man \fsn + \Msn \Man \fsd
] + i \gp [\Mpn^2 \fsd + 2 \Mpn \Msd \fpn] \exp(-i \Delta k z) \nn \\
& & + \srt \opn \exp[i (k_p -k_a)z] \hat{d}_{\Omega}(z)
 - \frac{\alpha_a}{2} \fan + \sqrt{\alpha_a}\hat{v}_a(z), \label{ffa} \\
\frac{{\rm d}\fsn}{{\rm d}z} &=& i [\gz + \gn][|\Mpn|^2 \fsn + \Mpd
 \Msn \fpn + \Mpn \Msn \fpd] + i \gz [2|\Msn|^2 \fsn + \Msn^2 \fsd] \nn \\
 & & + i[\gz + \gnn][|\Man|^2 \fsn + \Mad \Msn \fan + \Man \Msn \fad] + i \gn [\Mpn^2 \fad + 2 \Mpn \Mad \fpn]
\exp(-i \Delta k z) \nn \\ & & -\srt \opn \exp[i (k_p -k_s)z]
\hat{d}_{\Omega}^{\dagger}(z)  - \frac{\alpha_s}{2} \fsn +
\sqrt{\alpha_s}\hat{v}_s(z). \label{ffs}
\end{eqnarray}
When the amplifier is operating in the unsaturated regime, the pump
is depleted to a negligible degree. Thus one may neglect the four
wave-mixing terms in Eq. (\ref{fMp}). We next make the strong pump
approximation, i.e., $ |\Mpn|^2 \gg |\Man|^2, |\Msn|^2 $ , which is
valid at the input of the fiber under typically used operating
conditions and remains valid when there is essentially no pump
depletion.  Under these conditions, we note that the cross-phase
modulation of the pump wave due to the Stokes and anti-Stokes fields
is negligible.  We thus obtain the final form of the mean-field
equations:
\begin{eqnarray}
\frac{{\rm{d}} \Mpn }{{\rm{ d}}z} &=& i \, \gz \, \mps \Mpn -
\frac{\alpha_p}{2}\bar{A}_p, \label{meanfieldp_two} \\
\frac{{\rm{d}} \Man }{{\rm{d}}z} &=& i \, ( \gz + \gp ) \, \mps \Man
+ i \gp \Mpn^2 \Msd \phimn  -
\frac{\alpha_a}{2}\Man \label{meanfielda_two},\\
\frac{{\rm{d}} \Msn }{{\rm{d}}z} &=& i \, ( \gz + \gn ) \, \mps \Msn
+ i \gn \Mpn^2 \Mad \phimn  - \frac{\alpha_s}{2}\Msn.
\label{meanfields_two}
\end{eqnarray}
The solution of these mean-field equations [Eqs.
(\ref{meanfieldp}-\ref{meanfields})] were examined in the previous
section. Examining the fluctuation equations, we note that the
first-order fluctuation equations [Eqs. (\ref{ffp}-\ref{ffs})] are
linear in their quantum operators. The higher-order fluctuation
terms are nonlinear in the quantum-fluctuation operators, but their
contribution is negligible because they cannot contain terms
proportional to $\Mpn^2$.  We thus neglect the higher-order
fluctuation equations which are not reported here.

Similarly, those fluctuation terms in Eqs. (\ref{ffp}-\ref{ffs})
that do not contain two pump factors can be neglected, as they are
much weaker than the other terms.  We remark here that the
Langevin-noise terms due to the Raman effect in Eqs.
(\ref{ffa}-\ref{ffs}) will not be neglected.  This is justified
because comparison of the amplitude of the Langevin terms, $\srt
\opn$, in Eqs. (\ref{ffa}-\ref{ffs}) with the the amplitude of those
four-wave-mixing terms in Eqs. (\ref{ffa}-\ref{ffs}) that contain
only one pump term, for example $\gn \Mpn \Mad \fpn$  shows that the
Langevin term is much greater than the one-pump-field
four-wave-mixing terms. For example, consider the symmetrized
magnitude of the two terms:
\begin{eqnarray}
\sqrt{\frac{2 {\rm Im} \{  \gn \} |\Mpn|^2 [\langle
\hat{d}^{\dagger}_{\Omega} \hat{d}_{\Omega} +
 \hat{d}_{\Omega}  \hat{d}^{\dagger}_{\Omega}  \rangle] /2}{4 \gn^2 |\Man|^2 |\Mpn|^2 [\langle
\fpd \fpn + \fpn \fpd \rangle] /2} } = \sqrt{ \frac{ {\rm Im} \{ \gn
\} (\nth +1) }{2 |\gn|^2 \langle \hat{n}_a \rangle}}.
\end{eqnarray}
For an anti-Stokes amplitude of up to 10 mW (a flux of $ \hat{n}_a =
7.8 \times 10^{16}$ photons/s ) this ratio is greater than 25 for
pure silica and even better for materials with stronger Raman gains
(such as dispersion-shifted and highly nonlinear fibers). This means
that the Langevin noise term should be included.

One last simplification can be made to the pump equation, Eq.
(\ref{ffp}).  We simply note that all of the remaining noise terms
in Eq.~(\ref{ffp}) affect only the pump field and that under the
undepleted, strong-pump approximation, the Stokes and anti-Stokes
fields do not interact with the pump fluctuations to first degree.
This is the case if one assumes that the pump is, or nearly is, in a
coherent state. Thus, as we are only interested in the noise
introduced during the amplification process to the Stokes and
anti-Stokes fields, we neglect the pump fluctuations and replace Eq.
(\ref{ffp}) with
\begin{eqnarray}
\frac{{\rm d}\fpn}{{\rm d}z} = 0. \label{ffp_short} \end{eqnarray}
This allows one to treat the pump field fully classically.  To
summarize the steps taken, the undepleted, strong, coherent-state
pump approximation yields the following equations for the
first-order fluctuations:
\begin{eqnarray}
\frac{{\rm d}\fan}{{\rm d}z} &=& i [\gz + \gp]|\Mpn|^2 \fan + i \gp
\Mpn^2 \fsd \exp(-i \Delta k z) \nn
\\ & & + \srt \opn \exp[i (k_p -k_a)z] \hat{d}_{\Omega}(z)
 - \frac{\alpha_a}{2} \fan + \sqrt{\alpha_a}\hat{v}_a(z), \label{ffa_short} \\
\frac{{\rm d}\fsn}{{\rm d}z} &=& i [\gz + \gn]|\Mpn|^2 \fsn
 + i \gn \Mpn^2 \fad \exp(-i \Delta k z) \nn \\ & &  -\srt \opn \exp[i (k_p -k_s)z]
\hat{d}_{\Omega}^{\dagger}(z) - \frac{\alpha_s}{2} \fsn +
\sqrt{\alpha_s}\hat{v}_s(z). \label{ffs_short}
\end{eqnarray}
We note that Eqs. (\ref{ffp_short}-\ref{ffs_short}) can be combined
with Eqs. (\ref{meanfieldp_two}-\ref{meanfields_two}) by using Eq.
(\ref{lineq}) to yield:

\begin{eqnarray}
\frac{{\rm{d}} \Mpn }{{\rm{ d}}z} &=& i \, \gz \, \mps \Mpn -
\frac{\alpha_p}{2}\bar{A}_p,\\
\frac{{\rm{d}} \oan }{{\rm{d}}z} &=& i \, ( \gz + \gp)  \, \mps \oan + i \gp \Mpn^2 \osd \phimn -\frac{\alpha_a}{2} \hat{A}_a \nonumber \\
& & +\sqrt{2 {\rm Im} \{ \gp \} } \bar{A}_p\exp[i (k_p - k_a)z] \hat{d}_{\Omega}(z)  + \sqrt{\alpha_a}\hat{v}_a(z), \label{ch3eq4} \\
 \frac{{\rm{d}} \osn }{{\rm{d}}z} &=& i \, ( \gz + \gn) \, \mps
\osn + i \gn \Mpn^2 \oad
\phimn - \frac{\alpha_s}{2} \osn \nonumber \\
& &  -\sqrt{2 {\rm Im} \{ \gp \} } \bar{A}_p\exp[i (k_p -
k_s)z]\hat{d}_{\Omega}^{\dagger}(z) + \sqrt{\alpha_s}\hat{v}_s(z).
\label{ch3eq5}
\end{eqnarray}
This final form of the equations allows one to evolve the mean
fields of all three frequencies and the fluctuations of the Stokes
and anti-Stokes wavelengths simultaneously.}

The solution of Eqs.~(\ref{ch3eq4}) and (\ref{ch3eq5}) is
\begin{eqnarray}
\hat{A}_a(L) &=& \mu_a(0,L) \hat{A}_a(0) + \nu_a(0,L)
\hat{A}_s^{\dagger}(0) + \nonumber \\
& & \sqrt{2 {\rm Im} \{ \gp \} } \int_0^L {\rm d}z \, \bar{A}_p(z)
\exp[i(k_p-k_a)z]
\left[\mu_a(z,L)-\nu_a(z,L)\right]\hat{d}_{\Omega}(z),
\nonumber \\
& & +  \int_0^L {\rm d}z \, [ \sqrt{\alpha_a} \mu_a(z,L) \hat{v}_a(z) + \sqrt{\alpha_s} \nu_a(z,L) \hat{v}_s^{\dagger}(z)] \label{ch3eq12} \\
\hat{A}_s(L) &=& \mu_s(0,L) \hat{A}_s(0) + \nu_s(0,L)
\hat{A}_a^{\dagger}(0) +   \nonumber \\
& & \sqrt{2 {\rm Im} \{ \gp \} } \int_0^L {\rm d}z \, \bar{A}_p(z)
\exp[i(k_p - k_s)z] \left[ -\mu_s(z,L)+\nu_s(z,L)\right]
\hat{d}_{\Omega}^{\dagger}(z). \nonumber \\
& &+  \int_0^L {\rm d}z \,  [\sqrt{\alpha_s} \mu_s(z,L)
\hat{v}_s(z) + \sqrt{\alpha_a} \nu_s(z,L) \hat{v}_a^{\dagger}(z)]
\label{ch3eq13}
\end{eqnarray}

In the notation used in this paper, the functions $\mu_j(z,L)$ and
$\nu_j(L,z)$ denote evolution from a point $z$ in the fiber to the
end of the fiber ($L$) where the intensity and phase of the pump
at point $z$ must be used.

In this section, we have presented a thorough derivation of the
input-output mode transformations that govern a  $\chi^{(3)}$
parametric amplifier.  In the following two sections, we use these
input-output mode transformations to obtain the noise figure of
$\chi^{(3)}$ phase-sensitive parametric amplifiers and the squeezing
parameter for quadrature squeezing.

\section{Noise Figure of phase-sensitive amplification}
In this section we discuss the noise figure of phase-sensitive
amplification, which is defined as \begin{eqnarray} {\rm
NF}=\frac{{\rm SNR}_{\rm in}}{{\rm SNR}_{\rm out}}. \label{snreq}
\end{eqnarray} We assume for our treatment here that the
difference between the Stokes and anti-Stokes frequencies exceeds
the bandwidth of the detector. Thus beat frequencies of these two
waves will not be detected and can be neglected. { The mean photon
flux at the Stokes and anti-Stokes wavelength is $\langle \hat{n}_j
\rangle = \langle \hat{A}_j^{\dagger}(0) \hat{A}_j(0) \rangle =
|\zeta_j|^2$, and is assumed to be in a coherent state.} In what
follows, we neglect the small frequency difference between
$\omega_a$ and $\omega_s$. Thus the input SNR can be written as
\begin{eqnarray}
{\rm SNR}_{\rm in}= \frac{(\langle \hat{n}_a \rangle + \langle
\hat{n}_s \rangle)^2}{\langle \Delta \hat{n}^2_a \rangle + \langle
\Delta \hat{n}^2_s \rangle} =\frac{ (|\zeta_a|^2 +
|\zeta_s|^2)^2}{|\zeta_a|^2 + |\zeta_s|^2}= |\zeta_a|^2 +
|\zeta_s|^2.
\end{eqnarray}
Calculating the output SNR in a similar way and plugging into
Eq.~(\ref{snreq}), the NF  for phase-sensitive amplification can
be expressed as
\begin{eqnarray}
{\rm NF}= \frac{(|\zeta_a|^2 + |\zeta_s|^2)( \langle \Delta
\hat{n}^2_{\rm PI} \rangle  +  \langle \Delta \hat{n}^2_{\rm PS}
\rangle )}{(P_a + P_s)^2}, \label{nfpsa}
\end{eqnarray}
where the mean output power at each wavelength, $P_a$ and $P_s$,
is
\begin{eqnarray}
P_a &=& |\mu_a|^2|\zeta_a|^2 + |\nu_a|^2|\zeta_s|^2 + (\zeta_a \zeta_s \mu_a \nu_a^* + {\rm c.c.})  ,\\
P_s &=& |\mu_s|^2|\zeta_s|^2 + |\nu_s|^2|\zeta_a|^2 + (\zeta_a
\zeta_s \mu_s \nu_s^* + {\rm c.c.}).
\end{eqnarray}
In Eq.~(\ref{nfpsa}), we have expressed the variance of the output
photocurrent as the sum of a phase-insensitive portion, $\langle
\Delta \hat{n}^2_{\rm PI} \rangle$, and a phase-sensitive portion,
$\langle \Delta \hat{n}^2_{\rm PS} \rangle$, which are calculated
to be
\begin{eqnarray}
\langle \Delta \hat{n}^2_{\rm PI} \rangle &=& P_a B_a + P_s B_s,
\\
\langle \Delta \hat{n}^2_{\rm PS} \rangle &=& 2Q^* B_1 + 2Q B_2
,
\end{eqnarray}
where the quantities
\begin{eqnarray}
B_j &=& |\mu_j|^2 + |\nu_j|^2 + (2 \nth+1) |r_j|^2 + |c_{j1}|^2 + |c_{j2}|^2, \,\,\, (j=a,s)\\
  Q &=& (\mu_a \zeta_a + \nu_a \zeta_s^*)(\mu_s \zeta_s + \nu_s
  \zeta_a^*), \\
  B_{1} &=& c_{x1} + r_{x}(\nth +
1) +\mu_a \nu_s, \\
B_{2} &=& c_{x2}^* + r_{x}^* \nth + \mu_s^* \nu_a^*,
\end{eqnarray}
have noise terms defined as follows:
\begin{eqnarray}
|r_{a}|^2 &=& 2 {\rm Im} \{ \gp \}  \int_0^L {\rm d}z \,
|\bar{A}_p(z)|^2 | \mu_a(z,L)-\nu_a(z,L)|^2, \\
|r_{s}|^2 &=& - 2 {\rm Im} \{ \gn \}  \int_0^L {\rm d}z \,
|\bar{A}_p(z)|^2 | \mu_s(z,L)-\nu_s(z,L)|^2, \\
|c_{a(s)1}|^2 &=& \int_0^L {\rm d}z \, \alpha_{a(s)} |\mu_{a(s)}(z,L)|^2, \\
|c_{a(s)2}|^2 &=& \int_0^L {\rm d}z \, \alpha_{s(a)}
|\nu_{a(s)}(z,L)|^2, \\
r_{x} &=& 2 {\rm Im} \{ \gp \} \int_0^L {\rm d}z \, \bar{A}_p^2(z) \exp(-i \Delta k z) \nonumber \\ & & \times [ \mu_a(z,L)-\nu_a(z,L)][-\mu_s(z,L)+\nu_s(z,L)],\\
c_{x1} &=& \alpha_a \int_0^L {\rm d}z \, \mu_a(z,L) \nu_s(z,L), \\
 c_{x2}^* &=& \alpha_s \int_0^L {\rm d}z \, \nu_a(z,L)^*
\mu_s(z,L)^*.
\end{eqnarray}

 In the above expressions, $|r_{a(s)}|^2$ represents the integrated amplified
noise at the anti-Stokes (Stokes) wavelength seeded by thermally
populated optical phonon modes that are coupled in by the Raman
process. The terms $|c_{a(s)1}|^2$ represent integrated amplified
noise at the anti-Stokes (Stokes) wavelength seeded by vacuum
noise mixed in through distributed loss at the anti-Stokes
(Stokes) wavelength, while the terms $|c_{a(s)2}|^2$ represent
amplified noise at the anti-Stokes (Stokes) wavelength seeded by
vacuum noise mixed in through distributed loss at the Stokes
(anti-Stokes) wavelength.

In addition, the phase-sensitive terms $\mu_a \nu_s$ and $\mu_s^*
\nu_a^*$ represent amplified phase-sensitive noise seeded by the
vacuum noise at the anti-Stokes and Stokes wavelengths. The quantity
$r_{x}$ represents amplified phase-sensitive noise seeded by the
thermal-phonon fields due the Raman effect, and $c_{x1}$ and
$c_{x2}$ represent the amplified phase-sensitive noise seeded by the
vacuum noise due to distributed linear losses. Phase-sensitive noise
is present when the photocurrent variance with both Stokes and
anti-Stokes waves impinging on a detector is different from the sum
of the individual noise variances of the Stokes and anti-Stokes
frequencies.

\subsection{Degenerate Limit}

By taking the limiting value of the NF as $\Omega \rightarrow 0$,
we find the NF performance of a fully degenerate FOPA. We find
this limiting value of the NF by expanding the anti-symmetric
imaginary part of $\gp$ in a Taylor seris and expanding the
exponential in $n_{th}=\{ \exp[\hbar \Omega/(kT)]-1 \}^{-1}$
before allowing $\Omega \rightarrow 0$. We also use the fact that
in this limit, $\Delta k$ also approaches $0$ and the optimum
power splitting ratio approaches $0.5$.  This NF limit is:
\begin{eqnarray}
{\rm NF}_{\rm PSA, \, \Omega \rightarrow 0}= 1 + \frac{ \frac{4 k T
\bar{\gamma}_i^{'}(0)}{\hbar \gz} \left[ 1-\frac{\phi_{\rm
NL}}{\sqrt{1+\phi_{\rm NL}^2}} \right] }{1 + 2 \phi_{\rm NL}^2 + 2
\phi_{\rm NL} \sqrt{1 + \phi_{\rm NL}^2}},
\end{eqnarray}
where $\phi_{\rm NL}= \gz |\bar{A}_p|^2 L$ is the nonlinear phase
shift and $\bar{\gamma}_i^{'}(0)$ is the slope of the imaginary part
of $\gp$ as $\Omega \rightarrow 0$.  We observe that the PSA noise
figure for $\Omega \rightarrow 0$ increases to a maximum of slightly
more than $0$ dB and then decreases again and approaches $0$ dB in
the high-gain limit. This unusual behaviour of decreasing noise
figure vs. nonlinear phase shift is due to the relative scaling of
the Raman and FWM processes when $\Delta k =0$ and fact that the
mean Raman gain of the Stokes and anti-Stokes frequencies vanishes.
The total Raman noise scales linearly whereas the two-frequency
signal undergoes quadratic gain.  Consequently, the Raman noise
being created at the Stokes and anti-Stokes frequencies does not
grow.

\subsection{Results}

In Fig.~\ref{gainscan}, we plot the noise figure vs. PSA gain for
several values of detuning for a typical highly-nonlinear fiber with
a loss coefficient corresponding to $0.75$ dB$/$km. The plots show
that for detunings achievable by use of electrooptic elements (40
GHz detuning, phase-matched, { solid} curve), the results are almost
exactly the same as would be achieved in the limit of zero detuning
({ dotted}). For these simulations, we use realistic values for
$\bar{\gamma}(0)$ and for the distributed loss. We have additionally
assumed that the highly-nonlinear fiber has the same ratio of ${\rm
Im}\{ \bar{\gamma}(\Omega) \}$ to ${\rm Re}\{ \bar{\gamma}(\Omega)
\}$ as the standard dispersion-shifted fiber, i.e., the two have the
same germanium content.  {Unsurprisingly, we { see by comparing the
thick curves (lossy) to thin curves (lossless), that the distributed
loss increases the noise figure.} We aslo see that as the Raman-gain
coefficient decreases, the noise figure improves.

Interestingly, unlike the PIA case, Fig.~\ref{gainscan} shows that
the PSA noise figure is greater than $0$ dB as the gain approaches
$0$ dB. This occurs because the Raman gain and loss processes
dominate in the early parts of the amplifier (Raman gain and loss
are linear in the early parts of the amplifier while the
four-wave-mixing gain is quadratic), adding noise to both
frequencies, while the mean field undergoes no net gain due to the
Raman loss at one frequency and the Raman gain at the other.

In Fig.~\ref{nfspectrum}, we show the gain and noise-figure spectrum
for a $4$ km fiber with fiber parameters as described in the
caption.  This plot shows that the increasing Raman-gain coefficient
with detuning causes an increasing noise figure. These results show
that for realistic optical fibers, a FOPA operated as a PSA can
achieve a noise figure below $1$ dB for detunings up to $1$ THz.

\section{Nondegenerate quadrature squeezing}
When no light is injected into the FOPA, the quantum correlations
between the Stokes and anti-Stokes modes imply the presence of
quadrature squeezing, which is measurable by homodyne detection with
a two-frequency local oscillator (LO).

The difference current of the homodyne detector may be written as
\begin{eqnarray} \hat{I} = \hat{b}^{\dagger}_a\hat{q}_a + \hat{b}^{\dagger}_s
\hat{q}_s + {\rm H.c.}, \label{homodyne_current} \end{eqnarray}
where H.c. stands for the Hermitian conjugate of the first two
terms; $\hat{q}_a$ and $\hat{q}_s$ are the annihilation operators
corresponding to the anti-Stokes and Stokes components of the LO
beams, which are in a coherent state { each with photon flux}
$|\alpha_{{\rm LO},j}|^2$,  relative intensity
\begin{eqnarray} y_j=\frac{|\alpha_{{\rm LO},j}|^2}{|\alpha_{{\rm LO},a}|^2+|\alpha_{{\rm
LO},s}|^2},
\label{homodyne_splitting} \end{eqnarray} and phase $\theta_j$ for
$j=a,s$.

The squeezing parameter is defined as the ratio of the photocurrent
variance with the pump on to the photocurrent variance with vacuum
input to the homodyne detector (i.e., pump off).

\subsection{Lossless fiber with Raman effect}
In this paper, we concentrate on squezzing results for a lossless
FOPA. This is because increased nonlinear drive will, theoretically
at least, overpower linear loss, leading to no hard limit on the
achievable squeezing. However, as the Raman effect at each $z$
scales with pump intensity as does the four-wave-mixing process, the
lossless case illustrates a fundamental limit on the achievable
squeezing. Using Eqs.~(\ref{homodyne_current}) and
(\ref{homodyne_splitting}), we obtain, after some simple algebra,
for the lossless Raman-active case:
\begin{eqnarray} S &=& \frac{ \langle \Delta \hat{I}^2
\rangle } {\langle \Delta \hat{I}^2 \rangle_{\rm vac }} \nonumber
\\ &=& [1+ 2(|\nu_a|^2 + |r_a|^2 \nth)]y_a +
    \{1+ 2[|\nu_s|^2 + |r_s|^2 (1+\nth)]\}y_s \nonumber \\
    & & + 2\{ [\mu_s \nu_a(1+ \nth)- \mu_a \nu_s \nth] \exp[-i(\theta_a +
    \theta_s)]+ {\rm c.c.} \} \sqrt{y_a y_s}.
    \label{squeezing} \end{eqnarray}
In order to produce the best squeezing, one must choose the best
phase and relative intensity.  Once again, these two choices are
independent.  In order to choose the LO phases, we note that the
third term in Eq.~(\ref{squeezing}) has a negative sign when
\begin{eqnarray} \theta_a + \theta_s= \pi + \arg[\mu_s \nu_a (1+\nth) - \mu_a \nu_s
\nth]. \end{eqnarray}  Since the total LO power is conserved
($y_a+y_s=1$), we may use this fact to eliminate $y_s$ in
Eq.~(\ref{squeezing}), which then becomes quadratic in $y_a$.
Maximizing the magnitude of the third term in Eq.~(\ref{squeezing})
as a function of $y_a$ yields the additional condition for maximal
squeezing,
\begin{eqnarray}
y_a=\frac{1}{2} \left[ 1 + \frac{|r_s|^2(1+\nth)-|r_a|^2
\nth}{\sqrt{4 |\mu_s \nu_a(1+\nth) - \mu_a \nu_s \nth|^2 +
[|r_s|^2(\nth + 1) - |r_a|^2 \nth ]^2} } \right]. \end{eqnarray}
Thus the power splitting of the LOs for maximal squeezing is
slightly different from $50 \%$ and also slightly different from
that for maximal classical deamplification.

Use of this optimal two-frequency LO yields the following optimal
squeezing result
\begin{eqnarray}
S_{\rm opt} &=& 1 + |\nu_a|^2 + |\nu_s|^2 + |r_a|^2 \nth + |r_s|^2
(\nth+1) \nonumber \\ & & - \sqrt{4|\mu_s \nu_a (1+\nth)-\mu_a \nu_s
\nth|^2 + [|r_s|^2(1+\nth)-|r_a|^2 \nth ]^2}.
\end{eqnarray}

In order to make a connection with previous work, we show that in
the limit of degenerate operation we reach the same result as
obtained by Shapiro\cite{Shapiro95}. By placing the pump at the
zero-dispersion wavelength of the fiber, our solutions are then
identical to those in a dispersionless fiber (assuming the
higher-order terms in an expansion of $\beta$ are negligible). We
thus use the expressions in Eqs.~(\ref{mua_dkz}) to (\ref{nus_dkz})
for $\mu_j$ and $\nu_j$. In order to evaluate the limit as the
detuning approaches zero, we expand the anti-symmetric imaginary
part of $\gp$ as an odd power series around $\Omega=0$ and take the
Taylor series expansion of the exponential in $\nth = 1/[ \exp(\hbar
\Omega / (k T))-1  ]$. Then taking the limit as $\Omega \rightarrow
0$, the squeezing approaches the limit derived by
Shapiro\cite{Shapiro95} for the fully degenerate case.  This limit
is
\begin{eqnarray}
S_{\rm opt}(\Omega \rightarrow 0) &=& 1+ 2 \phi_{\rm NL} \left[
\phi_{\rm NL} + \frac{2 k
T \gamma_i^{'}(0)}{\hbar \gz } \right] \nonumber \\
& & - 2 \phi_{\rm NL} \left \{ 1 + \left[ \phi_{\rm NL} + \frac{2 k
T \bar{\gamma}_i^{'}(0)}{\hbar \gz} \right]^2 \right \}^{1/2},
\label{shapiroresult}
\end{eqnarray}
where $\phi_{\rm NL}= \gz |\bar{A}_p|^2 L$ is the nonlinear phase
shift and $\bar{\gamma}_i^{'}(0)$ is the slope of the imaginary part
of $\gp$ as $\Omega \rightarrow 0$.

In Fig~\ref{lossless_squeezing}, the main features of this squeezing
theory are illustrated for a lossless fiber. First, it is clear that
the Raman effect degrades the achievable amount of squeezing. In
addition, it can be seen by comparing the dashed lines
(phase-matched), dash-dotted lines (partially phase-matched) and
other lines ($\Delta k=0$) that when the Raman effect is included,
phase-matching leads to worse squeezing instead of improved
squeezing predicted by an instantaneous nonlinearity model. Finally,
by comparing the Raman-included dashed and dash-dotted lines, we see
that when $\Delta k \neq 0$, precise unequal balancing of the
relative power splitting of the two LO frequencies is required. When
this optimum splitting is achieved, the squeezing can be seen to
approach a constant value. However, when $\Delta k=0$, the possible
amount of squeezing is not bounded, and squeezing asymtotically
scales as $1/ \phi_{\rm NL}$, which is explained by the quadratic
scaling of the four-wave-mixing process and the linear scaling of
the Raman process. The hard limit of constant squeezing with
increasing nonlinear drive in the phase-matched ($\Delta k \neq 0$
and $\kappa=0$) case occurs when the excess noise due to the Raman
effect balances the strength of the squeezing process.

\section{Conclusion}

In conclusion, we have presented a quantum theory of parametric
amplification in a $\chi^{(3)}$ nonlinear medium that includes the
noninstantaneous response of the nonlinearity and the effect of
distributed linear loss.  We have applied this theory to
nondegenerate phase-sensitive amplification and deamplification and
have found the input conditions for optimal amplification and
deamplification.  We have also found the input conditions for
operation at the minimum noise figure, which { for detunings $< 1$
THz} is predicted to be $<0.4$ dB in the high-gain limit for FOPAs
made from typical dispersion-shifted fibers. We anticipate that
nondegenerate phase-sensitive amplifiers will produce record
noise-figure performance, as they allow circumvention of GAWBS noise
that is present in the degenerate case.  We have also presented a
theory of non-degenerate squeezing and found the optimal
continuous-wave local oscillator for a lossless FOPA with
non-instantaneous nonlinear response.  Our results agree with the
limit previously found by Shapiro\cite{Shapiro95} as degeneracy is
reached. Away from degeneracy and with a nonzero linear
phase-mismatch, we have shown that optimal squeezing in a dispersive
fiber when $\Delta k \neq 0$ reaches a constant limit, unlike the
$1/\phi_{\rm NL}$ scaling that occurs when the linear phase-mismatch
vanishes.

\section{Acknowledgments}
This work was supported by the U.S. Army Research Office, under a
MURI grant DAAD19-00-1-0177.


\newcommand{\noopsort}[1]{} \newcommand{\printfirst}[2]{#1}
  \newcommand{\singleletter}[1]{#1} \newcommand{\switchargs}[2]{#2#1}

\begin{figure}[tbh]
  \centerline{\epsfxsize=5.5in \epsfbox{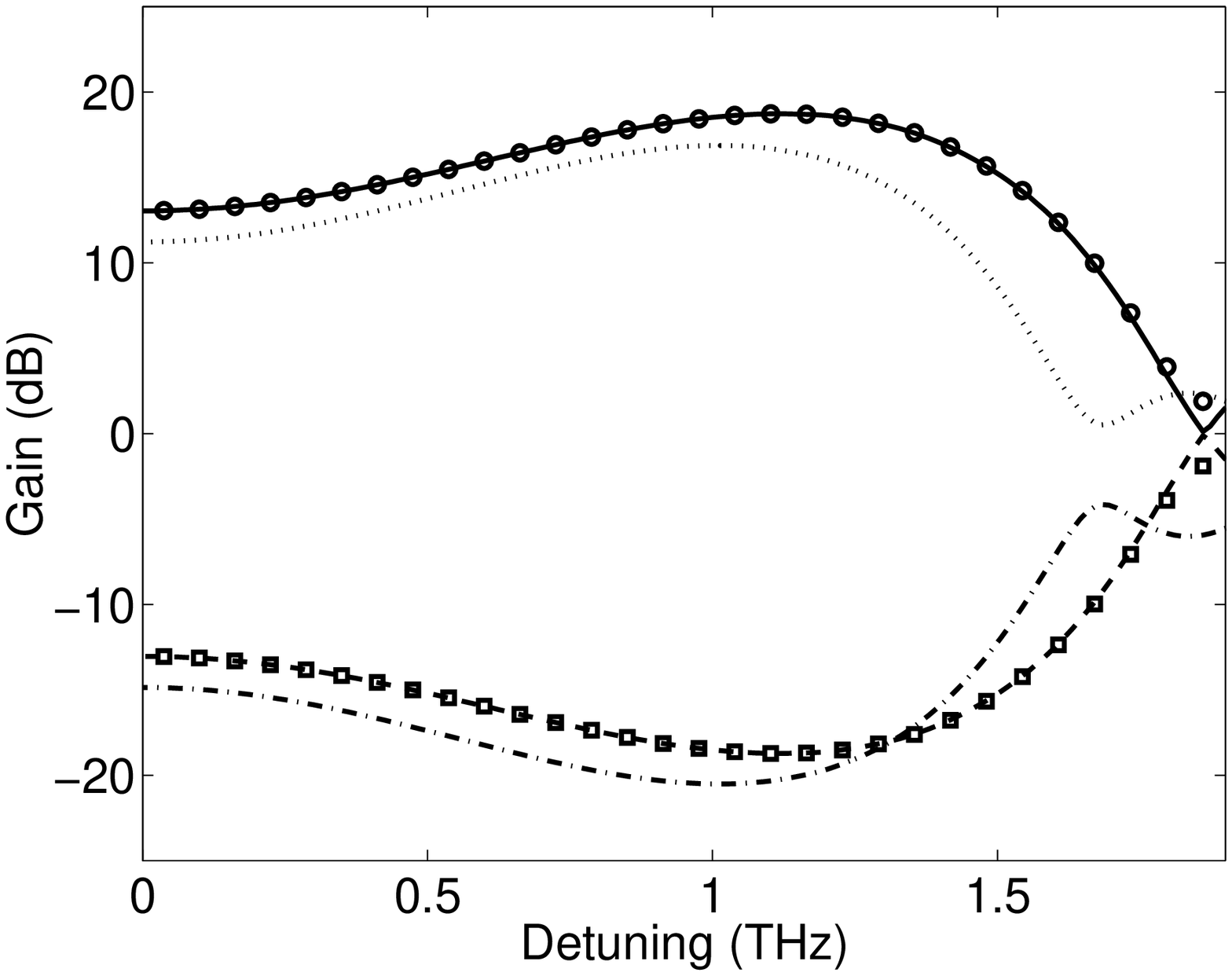}}
  \caption{{ Gain spectra vs. detuning for optimized PSAs made from (a) lossless
  fiber of
  length $L=3.63$ km and ${\rm Im} \{ \gamma_{\Omega} \}=0$
  (solid curve), (b) same as (a), but ${\rm Im} \{ \gamma_{\Omega} \}$
  calculated for DSF as explained in the text (circles), (c) $4.44$ km of DSF with $L_{\rm eff}=3.63$ km
  for
  $\alpha_a = \alpha_s = \alpha_p = 0.41$ dB$/$km and
  ${\rm Im} \{ \gamma_{\Omega} \}$ calculated for DSF as explained in the text
(dotted curve).  Gain spectra vs. detuning for optimized PSDs made
from (d) lossless
  fiber of
  length $L=3.63$ km and ${\rm Im} \{ \gamma_{\Omega} \}=0$
  (dashed curve), (b) same as (a), but ${\rm Im} \{ \gamma_{\Omega} \}$
  calculated for DSF as explained in the text (squares), (c) $4.44$ km of DSF with $L_{\rm eff}=3.63$ km
  for
  $\alpha_a = \alpha_s = \alpha_p = 0.41$ dB$/$km and
  ${\rm Im} \{ \gamma_{\Omega} \}$ calculated for DSF as explained in the text (dash-dotted curve). Input pump power is $0.33$
  Watts,  $\lambda_0=1551.16$ nm, pump wavelength is $1551.5$ nm,  and the dispersion slope is $57$ ps$/$(nm$^2$
  km).
  }} \label{psa_mean2}
  \end{figure}

 \begin{figure}[tbh]
  \centerline{\epsfxsize=5.5in \epsfbox{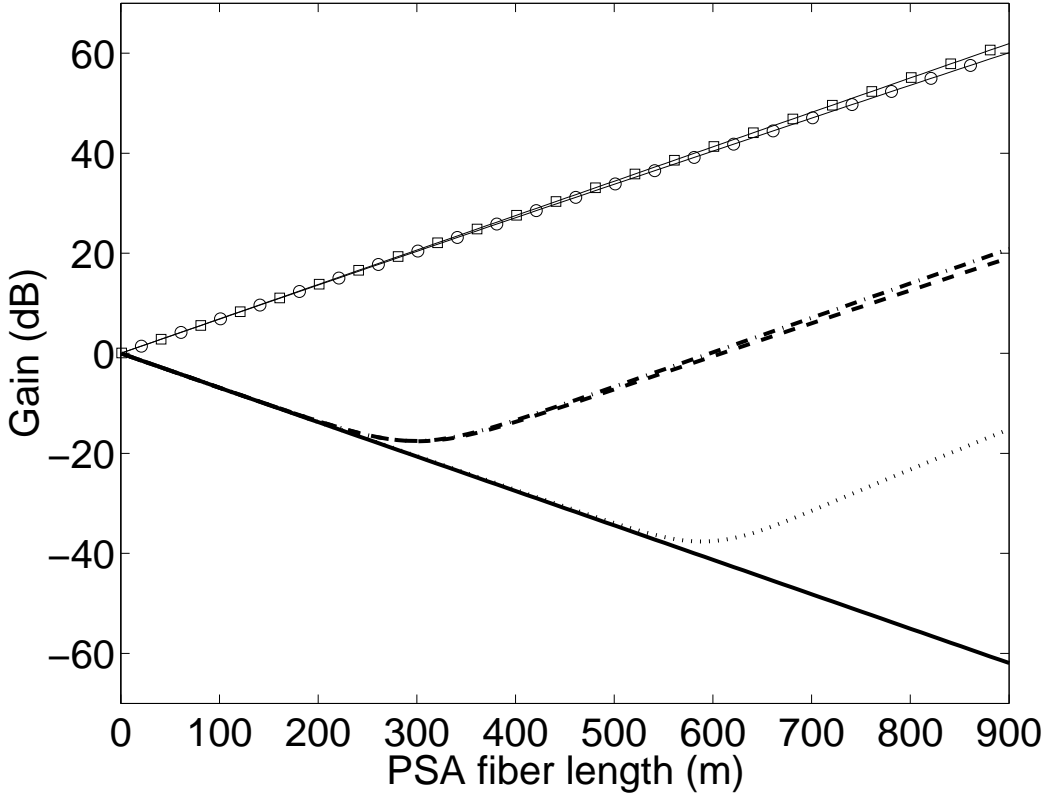}}
  \caption{Gain vs. fiber length for PSA made from DSF for  (a) phase-sensitive deamplification with optimum power splitting
  in lossless fiber
  ($\alpha_a=\alpha_s=\alpha_p=0$) (thick solid line), (b) phase-sensitive deamplification with optimum power splitting in a
  lossy fiber ($\alpha_a=\alpha_s=\alpha_p=0.25$ dB$/$km) (dotted
  line),
  (c) phase-sensitive deamplification in a lossless fiber with
  $|\zeta_a|^2=|\zeta_s|^2$ (dash-dotted line), (d)
  phase-sensitive deamplification in a lossy fiber with $|\zeta_a|^2=|\zeta_s|^2$ (dashed
  line), (e) phase-sensitive amplification in a lossless fiber
  with $|\zeta_a|^2=|\zeta_s|^2$ (squares) and optimum input power
  splitting (solid curve under the squares), and (f) phase-sensitive
  amplification in a lossy fiber with $|\zeta_a|^2=|\zeta_s|^2$ (circles) and optimum input power
  splitting (solid curve under the circles).  Input pump power is $4$
  Watts,  pump-signal detuning is $1$ THz, and phase-matching is
  achieved at the input [$\Delta k = -2 {\rm Re} \{\gamma_{\Omega} \} I_p(0)$].}
\label{psa_mean1}
  \end{figure}

\begin{figure}[h]
  \centerline{\epsfxsize=5.5in \epsfbox{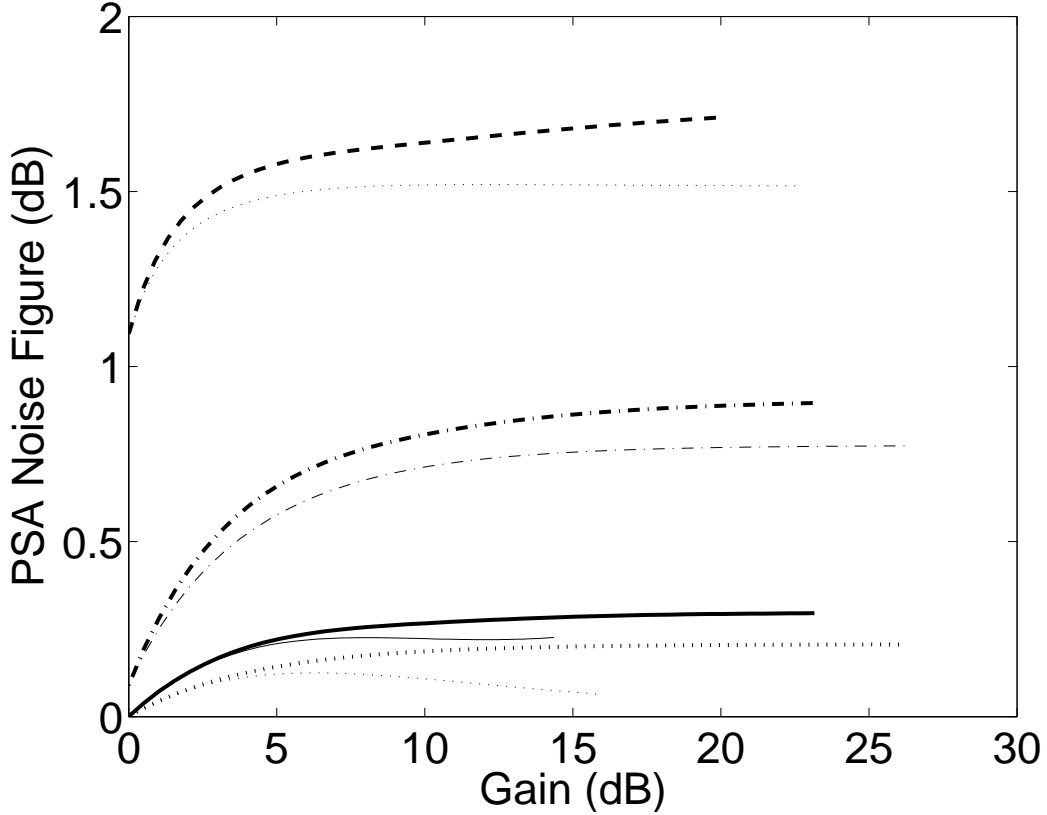}}
  \caption{  PSA noise figure vs. gain for various detunings for a highly-nonlinear fiber.
  For thick lines, fiber attenuation is $0.75$ dB/km at pump, Stokes, and anti-Stokes wavelengths; for thin lines,
  fiber is lossless.  $\Omega/2 \pi=13.8$ THz, dashed curves; $\Omega/2 \pi = 1.38$
  THz, dash-dotted curves; $\Omega/2 \pi = 40$ GHz, solid curves;
  and $\Omega/2 \pi = 0$ Hz, dotted curves.  Except for dotted curves, phase matching
  at the input [$\Delta k = -2 {\rm Re} \{ \gamma_{\Omega} \} I_p(0)$] is
  achieved.   For dotted curves, $\Delta k=0$.  The anti-Stokes/Stokes relative
  phase and power splitting at the input is for optimal classical gain.
  Initial pump power is $340$ mW, $\gamma(0)=9\times 10^{-3}$
  W$^{-1}$m$^{-1}$, peak imaginary part of $\gamma_{\Omega}$ is $3.5 \times
  10^{-3}$
  W$^{-1}$m$^{-1}$.
  Fiber length is $1$ km.} \label{gainscan}
  \end{figure}

\begin{figure}[h]
  \centerline{\epsfxsize=5.5in \epsfbox{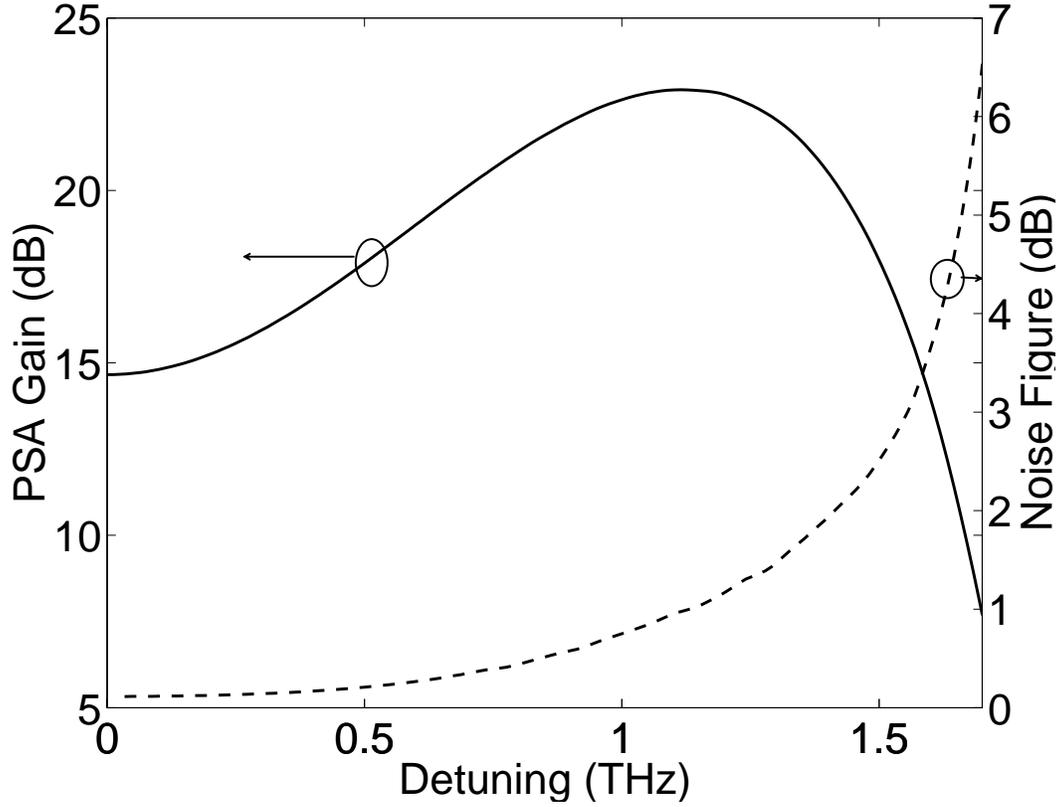}}
  \caption{  PSA gain and noise-figure spectrum vs. detuning.
  Anti-Stokes/Stokes relative phase and input-power splitting is for optimal classical gain.
  Initial pump power is $300$ mW, $\gamma_{0}=2 \times 10^{-3}$
  W$^{-1}$m$^{-1}$, peak imaginary part of $\gamma_{\Omega}$ is $0.75 \times
  10^{-3}$
  W$^{-1}$m$^{-1}$.  Attenuation is $0.41$ dB/km at the pump, Stokes, and anti-Stokes wavelengths.
  Fiber length is $4$ km.  $\lambda_0=1551.15$ nm, $\lambda_p=1555.5$ nm, and the dispersion slope is
  $57$ ps$/$nm$/$km$^2$ } \label{nfspectrum}
  \end{figure}

 \begin{figure}[h]
  \centerline{\epsfxsize=5.5in \epsfbox{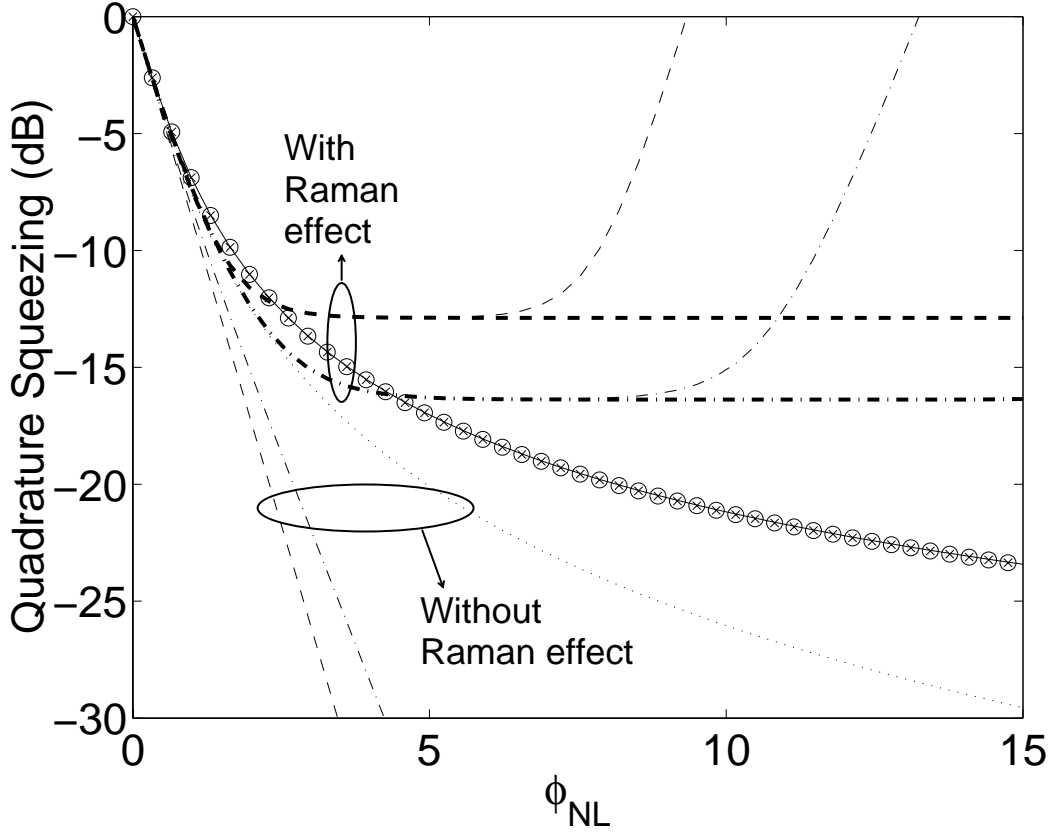}}
  \caption{  Squeezing vs. nonlinear phase shift in a lossless
  fiber.  Lower lines, without Raman effect.  Upper lines, with
  Raman effect.
  Dashed lines signify that phase-matching is
  achieved at the input [$\Delta k = -2 {\rm Re} \{ \gamma_{\Omega} \} |A_p|^2$].   The thick dashed line is for optimal
  LO power splitting, thin dashed lines for equal LO power splitting.
  Dash-dotted lines are for $ \Delta k = -(2/3) {\rm Re} \{ \gamma_{\Omega} \} |A_p|^2$. Thick dash-dotted line is for
  optimal LO power splitting, thin dashed lines for equal LO power splitting.  In all other lines, $\Delta k = 0$.
  Raman effect neglected, dotted line;  Raman effect included and
  equal LO power splitting, marked with x; Raman effect included
  and
  optimal LO power splitting, circles; CW limit of  Eq.~(\ref{shapiroresult})
  with
  $\bar{\gamma}_i^{'}(0) k T / \hbar \gz =0.026$, thin solid line.
  Pump-signal detuning is 40 GHz. }
\label{lossless_squeezing}
  \end{figure}
\end{document}